\documentclass[twocolumn,aps,prc,superscriptaddress,showpacs,floatfix]{revtex4}
\usepackage{amssymb}
\usepackage{amsmath}
\usepackage{graphicx}
\usepackage{longtable}
\usepackage[normalem]{ulem}
\usepackage[dvips]{color}

\renewcommand\sout{\bgroup \color{red} \ULdepth=-.5ex \ULset}

\begin{document}

\title{A phenomenological equation of state for isospin asymmetric nuclear
matter}
\author{CHEN LieWen}
\affiliation{Department of Physics, Shanghai Jiao Tong University, Shanghai 200240, China}
\affiliation{Center of Theoretical Nuclear Physics, National Laboratory of Heavy Ion
Accelerator, Lanzhou 730000, China}
\date{\today }

\begin{abstract}
A phenomenological momentum-independent (MID) model is constructed to
describe the equation of state (EOS) for isospin asymmetric nuclear matter,
especially the density dependence of the nuclear symmetry energy $E_{\text{%
\textrm{sym}}}(\rho )$. This model can reasonably describe the general
properties of the EOS for symmetric nuclear matter and the symmetry energy
predicted by both the sophisticated isospin and momentum dependent MDI model
and the Skyrme-Hartree-Fock approach. We find that there exists a nicely
linear correlation between $K_{\mathrm{sym}}$ and $L$ as well as between $%
J_{0}/K_{0} $ and $K_{0}$, where $L$ and $K_{\mathrm{sym}}$ represent,
respectively, the slope and curvature parameters of the symmetry energy at
the normal nuclear density $\rho _{0}$ while $K_{0}$ and $J_{0}$ are,
respectively, the incompressibility and the third-order derivative parameter
of symmetric nuclear matter at $\rho _{0}$. These correlations together with
the empirical constraints on $K_{0}$, $L$ and $E_{\text{\textrm{sym}}}(\rho
_{0}) $ lead to an estimation of $-477$ MeV $\leq K_{\mathrm{sat,2}}\leq
-241 $ MeV for the second-order isospin asymmetry expansion coefficient for
the incompressibility of asymmetric nuclear matter at the saturation point.
\end{abstract}

\pacs{Equation of state of nuclear matter, isospin, the symmetry
energy} \maketitle

\section{Introduction}

The study of the isospin degree of freedom in nuclear physics has
recently attracted much attention due to the establishment of many
radioactive beam facilities around the world. Besides the many
existing radioactive beam facilities and their upgrades, such as the
Cooling Storage Ring (CSR) facility at HIRFL in China \cite{CSR},
many more are being constructed or under planning, including the
Radioactive Ion Beam (RIB) Factory at RIKEN in Japan \cite{Yan07},
the FAIR/GSI in Germany \cite{FAIR}, SPIRAL2/GANIL in France
\cite{SPIRAL2}, and the Facility for Rare Isotope Beams (FRIB) in
the USA \cite{RIA}. These new facilities offer the possibility to
study the properties of nuclear matter or nuclei under the extreme
condition of large isospin asymmetry. The ultimate goal of such
study is to extract information on the isospin dependence of
in-medium nuclear effective interactions as well as the equation of
state (EOS) of isospin asymmetric nuclear matter, particularly its
isospin-dependent term or the density dependence of the nuclear
symmetry energy. This knowledge, especially the latter, is important
for understanding not only the structure of radioactive nuclei, the
reaction dynamics induced by rare isotopes, and the liquid-gas phase
transition in asymmetric nuclear matter, but also many critical
issues in astrophysics
\cite{LiBA98,LiBA01b,Dan02a,Lat00,Lat01,Lat04,Bar05,Ste05a,CKLY07,LCK08}.

The EOS of nuclear matter is one of fundamental questions in nuclear
physics. For symmetric nuclear matter, the EOS\ is relatively
well-determined after about more than $30$ years of studies in the nuclear
physics community. The incompressibility of symmetric nuclear matter at its
saturation density $\rho _{0}$ has been determined to be $240\pm 20$ MeV
from the nuclear giant monopole resonances (GMR) \cite%
{You99,Shl06,LiT07,Gar07,Col09} and the EOS at densities of $2\rho _{0}<\rho
<5\rho _{0}$ has also been constrained by measurements of collective flows
in nucleus-nucleus collisions \cite{Dan02a} and of subthreshold kaon
production \cite{Aic85,Fuc06a} in relativistic nucleus--nucleus collisions.
On the other hand, for asymmetric nuclear matter, the EOS, especially the
density dependence of the nuclear symmetry energy, is largely unknown.
Although the nuclear symmetry energy at $\rho _{0}$ is known to be around $%
30 $ MeV from the empirical liquid-drop mass formula \cite{Mey66,Pom03}, its
values at other densities are poorly known \cite{LiBA98,LiBA01b}. Various
microscopic and phenomenological models, such as the relativistic
Dirac-Brueckner-Hartree-Fock (DBHF) \cite%
{Ulr97,Fuc04,Ma04,Sam05a,Fuc05,Fuc05b,Ron06} and the non-relativistic
Brueckner-Hartree-Fock (BHF) \cite{Bom91,Zuo05,LiZH06} approach, the
relativistic mean-field (RMF) model based on nucleon-meson interactions \cite%
{Ren02,Bar05,Men06,Che07}, and the non-relativistic mean-field model based
on Skyrme-like interactions \cite%
{Das03,LiBA04a,LiBA04c,Che04,Riz04,Beh05,Riz05,Che05b}, have been used to
study the isospin-dependent properties of asymmetric nuclear matter, such as
the nuclear symmetry energy, the nuclear symmetry potential, the
isospin-splitting of nucleon effective mass, etc., but the predicted results
vary widely. In fact, even the sign of the symmetry energy above $3\rho _{0}$
is uncertain \cite{Bom01,Xiao09}. The theoretical uncertainties are mainly
due to the lack of knowledge about the isospin dependence of in-medium
nuclear effective interactions and the limitations in the techniques for
solving the nuclear many-body problem.

In the present work, we construct a phenomenological momentum-independent
(MID) model which can reasonably describe the general properties of
symmetric nuclear matter and the symmetry energy predicted by both the
sophisticated isospin and momentum dependent MDI model and the
Skyrme-Hartree-Fock approach with different Skyrme forces. In particular,
the density functional of the symmetry energy constructed in the MID model
is shown to be very flexible and can mimic very different density behaviors
by varying only one parameter. We find that there exists a nicely linear
correlation between $K_{\mathrm{sym}}$ and $L$ as well as between $J_{0}/K_{0}$ and $%
K_{0}$, where $L$ and $K_{\mathrm{sym}}$ represent, respectively,
the slope and curvature parameters of the symmetry energy at the
normal nuclear density $\rho _{0}$ while $K_{0}$ and $J_{0}$ are,
respectively, the incompressibility and the third-order derivative
parameter of symmetric nuclear matter at $\rho _{0}$. These
correlations together with the empirical constraints on $K_{0}$, $L$
and $E_{\text{\textrm{sym}}}(\rho _{0}) $ lead to an estimate of
$-477$ MeV $\leq K_{\mathrm{sat,2}}\leq -241 $ MeV for the
second-order isospin asymmetry expansion coefficient for the
incompressibility of asymmetric nuclear matter at the saturation
point, which is presently largely uncertain and being heavily
discussed~\cite{LiT07,Gar07,Pie07,Sag07,Pie09,Col09}.

The paper is organized as follows. In Section \ref{theory}, we discuss
general properties of asymmetric nuclear matter, and then introduce the
momentum independent MID model. The results and discussions are presented in
Section \ref{results}. A summary is then given in Section \ref{summary}.

\section{Theoretical models and methods}

\label{theory}

\subsection{Equation of state of asymmetric nuclear matter}

The EOS of isospin asymmetric nuclear matter, given by its binding
energy per nucleon, can be expanded to $2$nd-order in isospin
asymmetry $\delta $ as
\begin{equation}
E(\rho ,\delta )=E_{0}(\rho )+E_{\mathrm{sym}}(\rho )\delta ^{2}+O(\delta
^{4}),  \label{EOSANM}
\end{equation}%
where $\rho =\rho _{n}+\rho _{p}$ is the baryon density with $\rho _{n}$ and
$\rho _{p}$ denoting the neutron and proton densities, respectively; $\delta
=(\rho _{n}-\rho _{p})/(\rho _{p}+\rho _{n})$ is the isospin asymmetry; $%
E_{0}(\rho )=E(\rho ,\delta =0)$ is the binding energy per nucleon in
symmetric nuclear matter, and the nuclear symmetry energy is expressed as
\begin{equation}
E_{\mathrm{sym}}(\rho )=\frac{1}{2!}\frac{\partial ^{2}E(\rho ,\delta )}{%
\partial \delta ^{2}}|_{\delta =0}.  \label{Esym}
\end{equation}%
The absence of odd-order terms in $\delta $ in Eq. (\ref{EOSANM}) is due to
the exchange symmetry between protons and neutrons in nuclear matter when
one neglects the Coulomb interaction and assumes the charge symmetry of
nuclear forces. The higher-order coefficients in $\delta $ are usually very
small and negligible, e.g., the magnitude of the $\delta ^{4}$ term at
normal nuclear density $\rho _{0}$ is estimated to be less than $1$ MeV in
microscopic many-body approaches \cite{Sie70,Sjo74,Lag81}. Neglecting the
contribution from higher-order terms in Eq. (\ref{EOSANM}) leads to the
well-known empirical parabolic law for the EOS of asymmetric nuclear matter,
which has been verified by all many-body theories to date, at least for
densities up to moderate values \cite{LCK08}. As a good approximation, the
density-dependent symmetry energy $E_{\mathrm{sym}}(\rho )$ can thus be
extracted from the parabolic approximation of $E_{\mathrm{sym}}(\rho
)\approx E(\rho ,\delta =1)-E(\rho ,\delta =0)$.

Around the normal nuclear density $\rho _{0}$, the binding energy
per nucleon in symmetric nuclear matter $E_{0}(\rho )$\ can be
expanded, e.g., up to $3$rd-order in density as
\begin{equation}
E_{0}(\rho )=E_{0}(\rho _{0})+\frac{K_{0}}{2!}\chi ^{2}+\frac{J_{0}}{3!}\chi
^{3}+O(\chi ^{4}),  \label{E0}
\end{equation}%
where $\chi $ is a dimensionless variable characterizing the deviations of
the density from the saturation density $\rho _{0}$ of the symmetric nuclear
matter and it is conventionally defined as $\chi =(\rho -\rho _{0})/3\rho
_{0}$. $E_{0}(\rho _{0})$ is the binding energy per nucleon in symmetric
nuclear matter at the saturation density $\rho _{0}$ and the other
coefficients can be calculated as

\begin{equation}
K_{0}=9\rho _{0}^{2}\frac{d^{2}E_{0}(\rho )}{d\rho ^{2}}|_{\rho =\rho
_{0}},J_{0}=27\rho _{0}^{3}\frac{d^{3}E_{0}(\rho )}{d\rho ^{3}}|_{\rho =\rho
_{0}}.
\end{equation}%
Obviously, there is no linear $\chi $ term in Eq. (\ref{E0})
according to the definition of the saturation density $\rho _{0}$.
$K_{0}$ is the incompressibility coefficient of symmetric nuclear
matter and it characterizes the curvature of $E_{0}(\rho )$ at $\rho
_{0}$. The coefficient $J_{0}$ corresponds to the third-order
derivative parameter of symmetric nuclear matter at $\rho _{0}$. In
the literature, people usually neglect the higher-order terms in Eq.
(\ref{E0}) and obtain the following
parabolic approximation to the EOS\ of symmetric nuclear matter:%
\begin{equation}
E_{0}(\rho )=E_{0}(\rho _{0})+\frac{K_{0}}{2}\chi ^{2}+O(\chi ^{3}).
\label{E0para}
\end{equation}

Similarly, around $\rho _{0}$, the nuclear symmetry energy $E_{\mathrm{sym}%
}(\rho )$\ can be expanded, e.g., up to $2$nd-order in density as
\begin{equation}
E_{\mathrm{sym}}(\rho )=E_{\mathrm{sym}}(\rho _{0})+L\chi +\frac{K_{\mathrm{%
sym}}}{2!}\chi ^{2}+O(\chi ^{3}),  \notag
\end{equation}%
where $L$ and $K_{\mathrm{sym}}$ are the slope parameter and curvature
parameter of the nuclear symmetry energy at $\rho _{0}$, i.e.,
\begin{equation}
L=3\rho _{0}\frac{dE_{\mathrm{sym}}(\rho )}{\partial \rho }|_{\rho =\rho
_{0}},K_{\mathrm{sym}}=9\rho _{0}^{2}\frac{d^{2}E_{\mathrm{sym}}(\rho )}{%
\partial \rho ^{2}}|_{\rho =\rho _{0}}.
\end{equation}%
The coefficients $L$ and $K_{\mathrm{sym}}$ characterize the density
dependence of the nuclear symmetry energy around normal nuclear density $%
\rho _{0}$, and thus carry important information on the properties of
nuclear symmetry energy at both high and low densities.

The incompressibility is an essential quantity for nuclear matter
and conventionally it is defined at the saturation density $\rho
_{\mathrm{sat}}$ for asymmetric nuclear matter where we have $P(\rho
,\delta )=0$ (the incompressibility coefficient at the saturation
density is called isobaric incompressibility coefficient in
\cite{Pra85}) and thus it can be expressed as
\begin{equation}
K_{\mathrm{sat}}(\delta )=9\rho _{\mathrm{sat}}^{2}\frac{\partial ^{2}E(\rho
,\delta )}{\partial \rho ^{2}}|_{\rho =\rho _{\mathrm{sat}}}.
\label{KsatDef}
\end{equation}%
For asymmetric nuclear matter, the isobaric incompressibility coefficient $%
K_{\mathrm{sat}}(\delta )$ can be expressed up to $2$nd-order in
$\delta $ as~\cite{Bla80}

\begin{equation}
K_{\mathrm{sat}}(\delta )=K_{0}+K_{\mathrm{sat,2}}\delta ^{2}+O(\delta ^{4}),
\label{Ksat}
\end{equation}%
with%
\begin{equation}
K_{\mathrm{sat,2}}=K_{\mathrm{sym}}-6L-\frac{J_{0}}{K_{0}}L.  \label{Ksat2}
\end{equation}%
The coefficient $K_{\mathrm{sat,2}}$ essentially reflects the isospin
dependence of the isobaric incompressibility of asymmetric nuclear matter.

If we use the parabolic approximation to EOS of symmetric nuclear matter,
i.e.,\ Eq. (\ref{E0para}), then the $K_{\mathrm{sat,2}}$ is reduced to
\begin{equation}
K_{\mathrm{asy}}=K_{\mathrm{sym}}-6L  \label{Kasy}
\end{equation}%
and this expression has been extensively used to characterize the isospin
dependence of the incompressibility of asymmetric nuclear matter in the
literature~\cite{Lop88,Bar02,Che05a,Dan09}. Obviously, we have

\begin{equation}
K_{\mathrm{sat,2}}=K_{\mathrm{asy}}-\frac{J_{0}}{K_{0}}L,  \label{Ksat2Kasy}
\end{equation}%
and thus the coefficient $K_{\mathrm{asy}}$ could be a good approximation to
$K_{\mathrm{sat,2}}$ if $J_{0}$ is negligible or the slope parameter of the
symmetry energy $L$ is very small.

It is believed that information on $K_{\mathrm{sat,2}}$ can in principle be
extracted experimentally by measuring the GMR in neutron-rich nuclei~\cite%
{Bla80}. Usually, one can define a finite nucleus incompressibility $%
K_{A}(N,Z)$ for a nucleus with $N$ neutrons and $Z$ protons ($A=N+Z$) by the
energy of GMR $E_{\mathrm{GMR}}$, i.e.,
\begin{equation}
E_{\mathrm{GMR}}=\sqrt{\frac{\hbar ^{2}K_{A}(N,Z)}{m\left\langle
r^{2}\right\rangle }},
\end{equation}%
where $m$ is the nucleon mass and $\left\langle r^{2}\right\rangle $ is the
mean square mass radius of the nucleus at ground state. Similar to the
semi-empirical mass formula, the finite nucleus incompressibility $%
K_{A}(N,Z) $ can be expanded as%
\begin{equation}
K_{A}(N,Z)=K_{0}+K_{\mathrm{surf}}A^{-1/3}+K_{\tau }\left( \frac{N-Z}{A}%
\right) ^{2}+K_{\mathrm{Coul}}\frac{Z^{2}}{A^{4/3}},
\end{equation}%
where $K_{0}$, $K_{\mathrm{surf}}$, $K_{\tau }$, and $K_{\mathrm{coul}}$
represent the volume, surface, symmetry, and Coulomb terms, respectively.
The $K_{\tau }$ parameter is usually thought to be equivalent to the $K_{%
\mathrm{sat,2}}$ parameter. It should be noted here that the $K_{\mathrm{%
sat,2}}$ parameter is theoretically a well-defined physical quantity
while the value of the $K_{\tau }$ parameter may depend on the
detailed truncations in the expansion similarly to the
semi-empirical mass formula. Earlier attempts
based on the above method have given widely different values for the $%
K_{\tau }$ parameter. For example, a value of $K_{\tau }=-320\pm 180$ MeV
with a large uncertainty was obtained in Ref. \cite{Sha88} from a systematic
study of the GMR in the isotopic chains of Sn and Sm. In this analysis, the
value of $K_{0}$ was found to be $300\pm 25$ MeV, which is somewhat larger
than the commonly accepted value of $240\pm 20$ MeV. In a later study, an
even less stringent constraint of $-566\pm 1350<K_{\tau }<139\pm 1617$ MeV
was extracted from the GMR of finite nuclei, depending on the mass region of
nuclei and the number of parameters used in parameterizing the
incompressibility of finite nuclei \cite{Shl93}. Most recently, a much
stringent constraint of $K_{\tau }=-550\pm 100$ MeV has been obtained in
Ref. \cite{LiT07,Gar07} from measurements of the isotopic dependence of the
GMR in even-A Sn isotopes.

\subsection{A phenomenological momentum-independent MID model}

In the present work, we will mainly use three models, i.e., the isospin and
momentum dependent MDI interaction~\cite{Das03}, the Hartree-Fock approach
based on Skyrme interactions, and a phenomenological momentum-independent
interaction (MID). The MDI interaction~\cite{Das03} is based on the
finite-range Gogny effective interaction and has been used extensively in
the literature~\cite{LCK08}. The SHF approach is a well-known mean-field
theory and has been extensively used in the literature for its simplicity. A
very useful feature of these models is that analytical expressions for many
interesting physical quantities in asymmetric nuclear matter at zero
temperature can be obtained. Here we only introduce the MID model\ and for
the MDI and SHF models, one can refer to, e.g., Refs.~\cite{Das03,Cha97}.

In the momentum-independent MID model, following the results from
SHF approach with the zero-range and momentum-independent Skyrme
interaction, the potential energy density
$V_{\text{\textrm{MID}}}(\rho ,\delta )$ of a cold symmetric nuclear
matter at total density $\rho $ and isospin asymmetry $\delta $ is
parametrized as
\begin{equation}
V_{\text{\textrm{MID}}}(\rho ,\delta )=\frac{\alpha }{2}\frac{\rho ^{2}}{%
\rho _{0}}+\frac{\beta }{\sigma +1}\frac{\rho ^{\sigma +1}}{{\rho _{0}}%
^{\sigma }}+\rho E_{\text{\textrm{sym}}}^{pot}({\rho })\delta ^{2}.
\end{equation}%
In the MID model, the $4$th-order and higher-order nuclear symmetry energy
are not included and we assume they can be negligible. The parameters $%
\alpha $, $\beta $ and $\sigma $ are determined by the binding energy per
nucleon $E_{0}(\rho _{0})=-16$ MeV and the incompressibility $K_{0}$ at the
saturation density $\rho _{0}=0.16$ fm$^{-3}$
\begin{eqnarray}
\alpha &=&-29.47-46.74\frac{K_{0}+44.21}{K_{0}-166.11}\text{ (MeV),} \\
\beta &=&23.37\frac{K_{0}+254.53}{K_{0}-166.11}\text{ (MeV),} \\
\sigma &=&\frac{K_{0}+44.21}{210.32},
\end{eqnarray}%
where the unit of $K_{0}$ is MeV.

For the potential part of the symmetry energy $E_{\text{\textrm{sym}}}^{pot}(%
{\rho })$ in the MID model, it is parametrized as%
\begin{equation}
E_{\text{\textrm{sym}}}^{pot}({\rho })=E_{\text{\textrm{sym}}}^{pot}({\rho
_{0}})(1-y)\frac{{\rho }}{{\rho _{0}}}+yE_{\text{\textrm{sym}}}^{pot}({\rho
_{0}})\left( \frac{{\rho }}{{\rho _{0}}}\right) ^{\gamma _{\mathrm{sym}}}
\label{EsymPotMID}
\end{equation}%
with $E_{\text{\textrm{sym}}}^{pot}({\rho _{0}})=E_{\text{\textrm{sym}}}({%
\rho _{0}})-E_{\text{\textrm{sym}}}^{kin}({\rho _{0}})=17.7$ MeV following $%
E_{\text{\textrm{sym}}}({\rho _{0}})=30$ MeV and $E_{\text{\textrm{sym}}%
}^{kin}({\rho _{0}})=\frac{\hbar ^{2}}{6m}\left( \frac{3\pi ^{2}}{2}{\rho
_{0}}\right) ^{2/3}=12.3$ MeV. The default value of the $\gamma _{\mathrm{sym%
}}$ parameter is taken to be $4/3$ in the MID model following the $E_{\text{%
\textrm{sym}}}({\rho })$ in the MDI interaction (we will see how the $\gamma
_{\mathrm{sym}}$ parameter affects the symmetry energy in the following).
Similarly to the $x$ parameter introduced in the MDI interaction \cite{Che05a}%
, the dimensionless $y$ parameter is introduced to mimic various $E_{\mathrm{%
sym}}(\rho )$ predicted by different microscopic and/or phenomenological
many-body theories for a fixed $\gamma _{\mathrm{sym}}$ parameter. As we
will show later, for $\gamma _{\mathrm{sym}}=4/3$, adjusting the $y$ value
can nicely reproduce the $E_{\mathrm{sym}}(\rho )$ in the MDI interaction
with $x=-1$, $0$, and $1$.

In the MID model, the EOS of symmetric nuclear matter can thus be written as

\begin{equation}
E_{0}(\rho )=\frac{3\hbar ^{2}}{10m}\left( \frac{3\pi ^{2}}{2}\rho \right)
^{2/3}+\frac{\alpha }{2}\frac{\rho }{\rho _{0}}+\frac{\beta }{\sigma +1}%
\left( \frac{\rho }{{\rho _{0}}}\right) ^{\sigma },  \label{E0MID}
\end{equation}%
and the symmetry energy can be expressed as

\begin{eqnarray}
E_{\text{\textrm{sym}}}(\rho ) &=&\frac{\hbar ^{2}}{6m}\left( \frac{3\pi ^{2}%
}{2}\rho \right) ^{2/3}  \notag \\
&&+[E_{\text{\textrm{sym}}}({\rho _{0}})-E_{\text{\textrm{sym}}}^{kin}({\rho
_{0}})](1-y)\frac{{\rho }}{{\rho _{0}}}  \notag \\
&&+y[E_{\text{\textrm{sym}}}({\rho _{0}})-E_{\text{\textrm{sym}}}^{kin}({%
\rho _{0}})]\left( \frac{{\rho }}{{\rho _{0}}}\right) ^{\gamma _{\mathrm{sym}%
}}  \label{EsymMID}
\end{eqnarray}%
which leads to%
\begin{eqnarray}
L &=&2E_{\text{\textrm{sym}}}^{kin}({\rho _{0}})+3\left[ E_{\text{\textrm{sym%
}}}({\rho _{0}})-E_{\text{\textrm{sym}}}^{kin}({\rho _{0}})\right]  \notag \\
&&+3y(\gamma _{\mathrm{sym}}-1)\left[ E_{\text{\textrm{sym}}}({\rho _{0}}%
)-E_{\text{\textrm{sym}}}^{kin}({\rho _{0}})\right]  \label{LMID} \\
K_{\text{\textrm{sym}}} &=&9y\gamma _{\mathrm{sym}}(\gamma _{\mathrm{sym}}-1)%
\left[ E_{\text{\textrm{sym}}}({\rho _{0}})-E_{\text{\textrm{sym}}}^{kin}({%
\rho _{0}})\right]  \notag \\
&&-2E_{\text{\textrm{sym}}}^{kin}({\rho _{0}}).  \label{KsymMID}
\end{eqnarray}

\section{Results and discussions}

\label{results}

\subsection{The nuclear symmetry energy and correlation between $L$ and $K_{%
\mathrm{sym}}$}

\begin{figure}[tbh]
\includegraphics[scale=0.8]{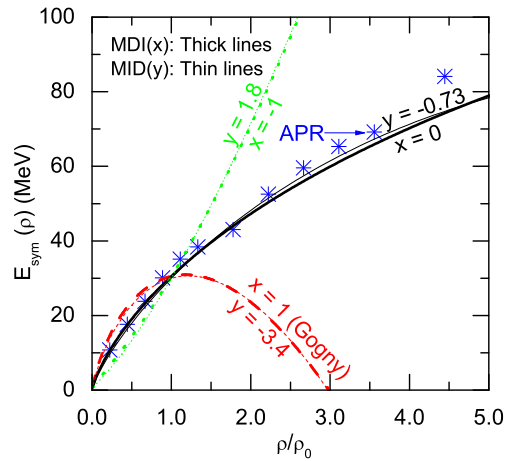}
\caption{{\protect\small (Color online) Density dependence of the symmetry
energy from the phenomenological MID interaction prediction with }$y=-3.4$%
{\protect\small , }$-0.73${\protect\small , and }$1.8${\protect\small . The
results from the MDI interaction with }$x=1${\protect\small , }$0$%
{\protect\small , and }$-1${\protect\small \ and the widely used APR
(Akmal-Pandharipande-Ravenhall) prediction \protect\cite{Akm98} are also
included for comparison.}}
\label{EsymRho}
\end{figure}
As mentioned above, in the MID model, the density dependence of the symmetry
energy can be adjusted by varying the $y$ parameter as shown in Eq. (\ref%
{EsymMID}). As an example, we show in Figure \ref{EsymRho} the
density
dependence of the symmetry energy from the MID interaction with $y=-3.4$, $%
-0.73$, and $1.8$. The corresponding results from the MDI interaction with $%
x=1$, $0$, and $-1$ as well as the widely used APR
(Akmal-Pandharipande-Ravenhall) prediction \cite{Akm98} are also
included for comparison. Indeed, one can see that the MID
interaction can give a nice description on the density dependence of
the symmetry energy predicted by the sophisticated MDI interaction
from the very soft ($x=1$) to the very stiff one ($x-1$).
Furthermore, it is seen that the APR prediction for the symmetry
energy at subsaturation densities lies right between that with $x=0$
and $-1$, and especially the symmetry energy with $x=0$ (and
$y=-0.73$) resembles very well the APR prediction up to about
$3.5\rho _{0}$. These features imply that the density functional of
the symmetry energy shown in Eq. (\ref{EsymMID}) is very flexible
and can give a quite general description for density dependence of
the symmetry energy.

The parameters $L$ and $K_{\mathrm{sym}}$ are determined by the density
dependence of the symmetry energy around $\rho _{0}$. In recent years,
significant progress has been made both experimentally and theoretically in
extracting the information on the symmetry energy at sub-saturation density
from heavy-ion reactions. Using the isospin and momentum-dependent IBUU04
transport model with in-medium NN cross sections, the isospin diffusion data
were found to be consistent with the symmetry energy from the MDI
interaction with $x$ between $0$ and $-1$, which can be parametrized by $E_{%
\mathrm{sym}}(\rho )\approx 31.6(\rho /\rho _{0})^{\gamma }$ with
$\gamma
=0.69-1.05$ at subnormal density ($\rho \leq \rho _{0}$) \cite%
{Tsa04,Che05a,LiBA05c,Che05b}, and has led to the extraction of $61$ MeV $%
\leq L\leq 111$ MeV and $-82$ MeV $\leq K_{\mathrm{sym}}\leq 101$ MeV~\cite%
{Tsa04,Che05a,LiBA05c,Che05b}. Using the Skyrme interactions consistent with
the EOS obtained from the MDI interaction with $x$ between $0$ and $-1$, the
neutron-skin thickness of heavy nuclei calculated within the Hartree-Fock
approach is consistent with available experimental data \cite{Che05b,Ste05b}
and also that from a relativistic mean-field model based on an accurately
calibrated parameter set that reproduces the GMR in $^{90}$Zr and $^{208}$Pb
as well as the isovector giant dipole resonance of $^{208}$Pb \cite{Tod05}.
The extracted symmetry energy further agrees with the symmetry energy $E_{%
\mathrm{sym}}(\rho )=31.6(\rho /\rho _{0})^{0.69}$ recently obtained from
the isoscaling analyses of isotope ratios in intermediate energy heavy ion
collisions \cite{She07}, which gives $L\approx 65$ MeV and $K_{\mathrm{sym}%
}\approx -61$ MeV. Furthermore, it is interesting to mention that the above
limited range of $E_{\mathrm{sym}}(\rho )$ at subsaturation density is
essentially consistent with the symmetry energy $E_{\mathrm{sym}}(\rho
)=12.5(\rho /\rho _{0})^{2/3}+17.6(\rho /\rho _{0})^{\gamma }$ with $\gamma
=0.4-1.05$, extracted very recently from analyses using the ImQMD (Improved
QMD) model which can reproduce both the isospin diffusion data and the
double neutron/proton ratio simultaneously \cite{Tsa09}. The symmetry energy
$E_{\mathrm{sym}}(\rho )=12.5(\rho /\rho _{0})^{2/3}+17.6(\rho /\rho
_{0})^{\gamma }$ with $\gamma =0.4-1.05$ thus leads to the constraints of $%
46 $ MeV $\leq L\leq 80$ MeV and $-82$ MeV $\leq K_{\mathrm{sym}}\leq -36$
MeV.
\begin{figure}[tbh]
\includegraphics[scale=0.8]{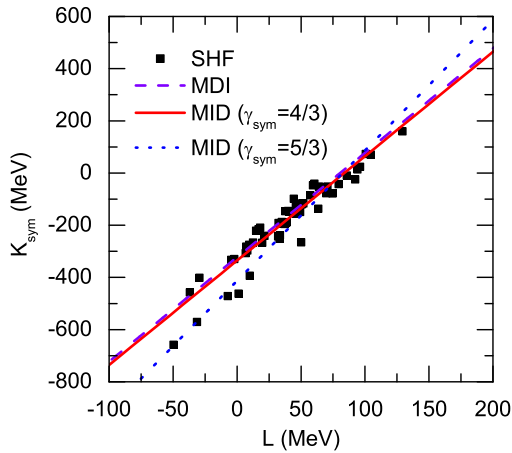}
\caption{{\protect\small (Color online) Correlation between\ }$K_{\mathrm{sym%
}}${\protect\small \ and }$L${\protect\small \ from the MID interaction with
}$\protect\gamma _{\mathrm{sym}}=4/3${\protect\small \ and }$5/3$%
{\protect\small , the MDI interaction and the SHF prediction with }$63$%
{\protect\small \ popular Skyrme forces.}}
\label{KsymL}
\end{figure}

It should be noted that all the above constraints on $L$ and $K_{\mathrm{sym}%
}$ are based on some unique energy density functionals and thus special
correlation between $L$ and $K_{\mathrm{sym}}$ has been implicitly assumed.
It is thus interesting to see if there exists a universal correlation
between $L$ and $K_{\mathrm{sym}}$. For the MDI interaction, the $L$ and $K_{%
\mathrm{sym}}$ both change linearly with the parameter $x$ and therefore
they are linearly correlated by varying the parameter $x$ \cite{Che05a,Xu09}%
. Similarly, for the MID interaction, one can see from Eq. (\ref{LMID}) and
Eq. (\ref{KsymMID}) that the $L$ and $K_{\mathrm{sym}}$ both change linearly
with the parameter $y$, and thus they are also linearly correlated by
varying the parameter $y$. In particular, we have
\begin{equation}
K_{\text{\textrm{sym}}}=3\gamma _{\mathrm{sym}}L+E_{\text{\textrm{sym}}%
}^{kin}({\rho _{0}})(3\gamma _{\mathrm{sym}}-2)-9\gamma _{\mathrm{sym}}E_{%
\text{\textrm{sym}}}({\rho _{0}}).  \label{KsymLMID}
\end{equation}%
Also the $L$ and $K_{\mathrm{sym}}$ are expected to be correlated
within the SHF energy density functional. Shown in Figure
\ref{KsymL} are the correlation
between\ $K_{\mathrm{sym}}$ and $L$ from the MID interaction with $\gamma _{%
\mathrm{sym}}=4/3$ and $5/3$ ($E_{\text{\textrm{sym}}}^{kin}({\rho _{0}}%
)=12.3$ MeV and $E_{\text{\textrm{sym}}}({\rho _{0}})=30$ MeV), the MDI
interaction and the SHF prediction with $63$ popular Skyrme forces. The $63$
Skyrme forces include the $51$ forces used in Ref. \cite{Xu09} and $12$ new
forces, i.e., Z, E$_{\sigma }$, E, Z$_{\sigma }$, Z$_{\sigma }^{\ast }$,
SkSC4, SI, SII, SIII, SIV, SV, and SVI. All these Skyrme forces predict the
saturation density and the symmetry energy satisfying $0.140$ fm$^{-3}<\rho
_{0}<0.165$ fm$^{-3}$ and $25$ MeV$<E_{sym}(\rho _{0})<37$ MeV, respectively.

It is interesting to see that the $K_{\mathrm{sym}}$ parameter indeed
displays approximately a linear correlation with the $L$ parameter for the
SHF prediction with the $63$ Skyrme forces and this linear correlation is
nicely reproduced by the MDI interaction and the MID interaction with $%
\gamma _{\mathrm{sym}}=4/3$. For the MID interaction, one can see from Eq. (%
\ref{KsymMID}) that the $\gamma _{\mathrm{sym}}$ parameter controls
the shape (slope) of the linear correlation between $L$ and
$K_{\mathrm{sym}}$. Furthermore, it is seen from Figure \ref{KsymL}
that there are a few Skyrme forces deviate from the linear
correlation\ obtained by the MDI interaction and the MID interaction
with $\gamma _{\mathrm{sym}}=4/3$. In order to consider the
uncertainty of the shape (slope) for the correlation between $L$
and $K_{\mathrm{sym}} $, we thus include the result with $\gamma _{\mathrm{%
sym}}=5/3$ for the MID interaction. The correlation between\ $K_{\mathrm{sym}%
}$ and $L$ from the SHF prediction with the $63$ Skyrme forces is nicely
consistent with that from the MID interaction with $\gamma _{\mathrm{sym}%
}=4/3$ and $5/3$. The linear correlation between\ $K_{\mathrm{sym}}$ and $L$
implies that one can obtain\ $K_{\mathrm{sym}}$ from $L$.

\subsection{Correlation between $J_{0}$ and $K_{0}$}

\begin{figure}[tbh]
\includegraphics[scale=0.8]{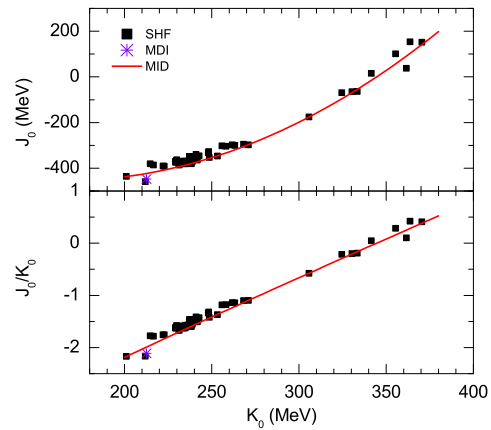}
\caption{{\protect\small (Color online) }$J_{0}${\protect\small \ and }$%
J_{0}/K_{0}${\protect\small \ as a function of }$K_{0}${\protect\small \
from the MID interaction, the MDI interaction and the SHF prediction with }$%
63${\protect\small \ popular Skyrme forces.}}
\label{J0K0abcSHFMDI}
\end{figure}

While $K_{0}$ has been relatively well determined, the $J_{0}$ parameter is
poorly known and actually there is no any experimental information on the $%
J_{0}$ parameter. In the MID model, from Eq. (\ref{E0MID}) one can easily
calculate the $J_{0}$ parameter as
\begin{eqnarray}
J_{0} &=&27\rho _{0}^{3}\frac{\partial ^{3}E_{0}(\rho )}{\partial ^{3}\rho }%
|_{\rho =\rho _{0}}  \notag \\
&=&\frac{1}{70.1}\left( K_{0}^{2}-332.2K_{0}-4243.2\right) \text{ (MeV),}
\label{J0}
\end{eqnarray}%
where the unit of $K_{0}$ is MeV. Therefore, in the MID model, the $J_{0}$
parameter is quadratically correlated with $K_{0}$. Shown in Figure \ref%
{J0K0abcSHFMDI} are $J_{0}$ and $J_{0}/K_{0}$ as functions of
$K_{0}$. Also included in Figure \ref{J0K0abcSHFMDI} are the
corresponding results from the MDI interaction and the SHF
prediction with the $63$ Skyrme forces. It is interesting to see
that the correlation between $J_{0}$ and $K_{0}$ is quite consistent
for the three different models, namely, the MID interaction, the MDI
interaction and the $63$ Skyrme forces in the SHF approach. In
particular, the $J_{0}/K_{0}$ displays approximately a linear
correlation with $K_{0}$. This linear correlation can be easily
understood from Eq. (\ref{J0}). On the r.h.s of Eq. (\ref{J0}), the
last term is very small compared with the first term and the second
term and thus one has $J_{0}\approx \frac{1}{70.1}\left(
K_{0}^{2}-332.2K_{0}\right) $, and then $J_{0}/K_{0}\approx
\frac{1}{70.1}\left( K_{0}-332.2\right) $ with the unit of $K_{0}$
being MeV. We note here that the correlation between $J_{0}$ and
$K_{0}$ obtained in the present work is also consistent with the
early finding by Pearson \cite{Pea91}. While there is no any
empirical constraint on the $J_{0}$ parameter, we assume in the
present work the correlation between $J_{0}$ and
$K_{0}$ from the MID interaction is valid and then we can obtain $%
J_{0}/K_{0} $ from the experimental constraint on $K_{0}$.

\subsection{Phenomenological MID model constraint on the $K_{\mathrm{sat,2}}$
parameter}

\begin{figure}[tbh]
\includegraphics[scale=0.75]{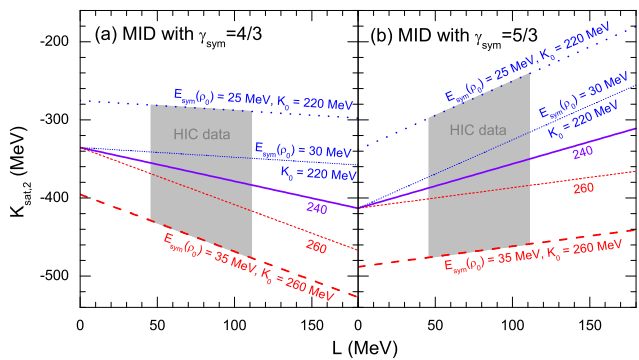}
\caption{{\protect\small (Color online) }$K_{\mathrm{sat,2}}${\protect\small %
\ as a function of }$L${\protect\small \ from the MID interaction with }$%
\protect\gamma _{\mathrm{sym}}=4/3${\protect\small \ (a) and }$5/3$%
{\protect\small \ (b) for different values of }$K_{0}${\protect\small \ and }%
$E_{\text{\textrm{sym}}}(\protect\rho _{0})${\protect\small . The shaded
region indicates constraints within MID interaction with }$220$%
{\protect\small \ MeV }$\leq K_{0}\leq 260${\protect\small \ MeV, }$25$%
{\protect\small \ MeV }$\leq E_{\text{\textrm{sym}}}(\protect\rho _{0})\leq
35${\protect\small \ MeV, and }$46${\protect\small \ MeV }$\leq L\leq 111$%
{\protect\small \ MeV limited by the heavy-ion collision data.}}
\label{Ksat2L}
\end{figure}
As shown in Eq. (\ref{Ksat2}), the $K_{\mathrm{sat,2}}$ parameter is
completely determined by $J_{0}/K_{0}$, $L$, and $K_{\mathrm{sym}}$.
Based on the correlations shown in Figure \ref{KsymL} and\ Figure
\ref{J0K0abcSHFMDI}, we can now extract information on the
$K_{\mathrm{sat,2}}$ parameter from the experimental constraints on
the $K_{0}$ parameter and the $L$ parameter within the MID model. As
pointed out previously, the value of $K_{0}$ has been relatively
well determined to be $240\pm 20$ MeV from the nuclear GMR
\cite{You99,Shl06,LiT07,Gar07,Col09}. The slope parameter $L$ has
been found to correlate linearly with the neutron-skin thickness of
heavy nuclei and thus can in principle be determined from measured
thickness of the neutron
skin of such nuclei \cite{Bro00,Hor01a,Typ01,Fur02,Kar02,Die03,Che05b,Ste05b}%
. Unfortunately, because of the large uncertainties in the experimental
measurements, this has not yet been possible so far. The proposed experiment
of parity-violating electron scattering from $^{208}$Pb, i.e., Parity Radius
Experiment (PREx) at the Jefferson Laboratory is expected to give an
independent and accurate measurement of its neutron skin thickness (within $%
0.05$ fm)~\cite{Hor01,Mic05}. On the other hand, as mentioned previously,
heavy-ion collisions, especially those induced by neutron-rich nuclei,
provide a unique tool to explore the density dependence of the symmetry
energy and thus the $L$ parameter.

In the MID model, from Eqs. (\ref{Ksat2}) and (\ref{KsymLMID}), we have

\begin{eqnarray}
K_{\mathrm{sat,2}} &=&-(\frac{J_{0}}{K_{0}}+6-3\gamma _{\mathrm{sym}%
})L+(3\gamma _{\mathrm{sym}}-2)E_{\text{\textrm{sym}}}^{kin}({\rho _{0}})
\notag \\
&&-9\gamma _{\mathrm{sym}}E_{\text{\textrm{sym}}}({\rho _{0}}).
\label{Ksat2MID}
\end{eqnarray}%
Shown in Figure \ref{Ksat2L} is $K_{\mathrm{sat,2}}$ as a function
of $L$ from the MID interaction with $\gamma _{\mathrm{sym}}=4/3$
(panel (a)) and $5/3$ (panel (b)) for $K_{0}=220$, $240$, and $260$
MeV. In Eq. (\ref{Ksat2MID}), $J_{0}$ can be obtained from Eq.
(\ref{J0}) for a fixed $K_{0}$ value. In the MID interaction,
$E_{\text{\textrm{sym}}}^{kin}({\rho _{0}})=12.3$ MeV and
$E_{\text{\textrm{sym}}}({\rho _{0}})=30$ MeV have been used as a
default. From Eq. (\ref{KsymLMID}), one can see that the correlation
of $K_{\mathrm{sym}}$ and $L$ also depends on
$E_{\text{\textrm{sym}}}({\rho _{0}})$. To consider the uncertainty
due to the $E_{\text{\textrm{sym}}}({\rho _{0}})$, we thus also
include in Figure \ref{Ksat2L} the results with $K_{0}=220$ MeV and
$E_{\text{\textrm{sym}}}({\rho _{0}})=25$ MeV as well as $K_{0}=260$
MeV and $E_{\text{\textrm{sym}}}({\rho _{0}})=35$ MeV, which
represent, respectively, the upper and lower boundaries for a fixed
$L$. The shaded region in Figure \ref{Ksat2L} further considers the
constrained $L$ values from heavy-ion collision data, namely, $46$
MeV $\leq L\leq 111$ MeV. The lower limit of $L=46$ MeV is obtained
from the lower boundary of the ImQMD analyses on the isospin
diffusion data and the double neutron/proton ratio \cite{Tsa09}
while the upper limit of $L=111$ MeV corresponds to the upper
boundary of $L$ from the IBUU04 transport model analysis on the
isospin
diffusion data \cite{Tsa04,Che05a,LiBA05c,Che05b}. The constraint $46$ MeV $%
\leq L\leq 111$ MeV is also consistent with the analyses of the pygmy dipole
resonances \cite{Kli07}, the giant dipole resonance (GDR) of $^{208}$Pb
analyzed with Skyrme forces \cite{Tri08}, the Thomas-Fermi model fitted very
precisely to binding energies of $1654$ nuclei~\cite{Mye96}, and the recent
neutron-skin analysis \cite{Cen09}. These empirically extracted values for $L
$ represent the best and most stringent phenomenological constraints
available so far on the nuclear symmetry energy at sub-saturation densities.

It is seen from Figure \ref{Ksat2L} that the $K_{\mathrm{sat,2}}$
decreases with increasing $L$ for $\gamma _{\mathrm{sym}}=4/3$ while
it increases with increasing $L$ for $\gamma _{\mathrm{sym}}=5/3$.
This feature can be easily
understood from Eq. (\ref{Ksat2MID}). For $\gamma _{\mathrm{sym}}=4/3$, Eq. (%
\ref{Ksat2MID}) is reduced to

\begin{equation}
K_{\mathrm{sat,2}}=-(\frac{J_{0}}{K_{0}}+2)L-12E_{\text{\textrm{sym}}}({\rho
_{0}})+24.6\text{ (MeV)}  \label{Ksat2L1}
\end{equation}%
while for $\gamma _{\mathrm{sym}}=5/3$, it is reduced to

\begin{equation}
K_{\mathrm{sat,2}}=-(\frac{J_{0}}{K_{0}}+1)L-15E_{\text{\textrm{sym}}}({\rho
_{0}})+36.9\text{ (MeV).}  \label{Ksat2L2}
\end{equation}%
For $K_{0}=240\pm 20$ MeV, $J_{0}/K_{0}$ can be found from Figure \ref%
{J0K0abcSHFMDI} (or Eq. (\ref{J0})) to be from about $-1.9$ to
$-1.3$. Therefore, $K_{\mathrm{sat,2}}$ decreases (increases) with
increasing $L$
for $\gamma _{\mathrm{sym}}=4/3$ ($5/3$) following Eq. (\ref{Ksat2L1}) (Eq. (%
\ref{Ksat2L2})).

An interesting feature observed from Figure \ref{Ksat2L} is that the $K_{%
\mathrm{sat,2}}$ parameter significantly depends on the symmetry
energy at the normal nuclear density $E_{\text{\textrm{sym}}}({\rho
_{0}})$. This can be seen more clearly from Eqs. (\ref{Ksat2L1}) and
(\ref{Ksat2L2}) which indicate that changing
$E_{\text{\textrm{sym}}}({\rho _{0}})$ by $5$ MeV leads to a
variation of $60-75$ MeV for $K_{\mathrm{sat,2}}$. This feature
indicates that an accurate determination of
$E_{\text{\textrm{sym}}}({\rho _{0}})$ is important for determining
the value of $K_{\mathrm{sat,2}}$. From
the shaded region indicated in Figure \ref{Ksat2L}, it is found that for $%
\gamma _{\mathrm{sym}}=4/3$, we have $-429$ MeV $\leq
K_{\mathrm{sat,2}}\leq -281$ MeV for $L=46$ MeV while $-477$ MeV
$\leq K_{\mathrm{sat,2}}\leq -289$
MeV for $L=111$ MeV. For $\gamma _{\mathrm{sym}}=5/3$, we have $-476$ MeV $%
\leq K_{\mathrm{sat,2}}\leq -298$ MeV for $L=46$ MeV while $-459$
MeV $\leq K_{\mathrm{sat,2}}\leq -241$ MeV for $L=111$ MeV. These
results indicate that within the MID model with the empirical
constraints of $K_{0}=240\pm 20$ MeV, $25$ MeV $\leq
E_{\text{\textrm{sym}}}(\rho _{0})\leq 35$ MeV,\ and $46$ MeV $\leq
L\leq 111$ MeV, the $K_{\mathrm{sat,2}}$\ parameter can be varied
from $-477$ MeV to $-241$ MeV.

As shown in Eq. (\ref{Ksat2Kasy}), the $K_{\mathrm{asy}}$ parameter
corresponds to the $K_{\mathrm{sat,2}}$ parameter when $J_{0}$ is zero,
i.e., the parabolic approximation to the EOS of symmetric nuclear matter Eq.
(\ref{E0para}) is valid. From the MID model, a vanishing $J_{0}$ corresponds
to a $K_{0}$ value of about $340$ MeV, which is significantly larger than
the empirical value of $240\pm 20$ MeV. In the MID model, we have $%
J_{0}/K_{0}\approx -1.6$ for $K_{0}=240$ MeV and thus $K_{\mathrm{sat,2}%
}\approx K_{\mathrm{asy}}+1.6L$ or $K_{\mathrm{asy}}\approx K_{\mathrm{sat,2}%
}-1.6L$. Therefore, the difference between $K_{\mathrm{asy}}$ and $K_{%
\mathrm{sat,2}}$ depends on $L$ with a larger\ $L$ value (stiffer symmetry
energy) leading to larger difference. At this point, it should be stressed
that the $K_{\mathrm{asy}}$ parameter is completely determined by the
density dependence of the symmetry energy regardless of the EOS\ of
symmetric nuclear matter. Based on the IBUU04 transport model analysis on
the isospin diffusion data \cite{Che05a,LiBA05c}, a value of $K_{\mathrm{asy}%
}=-500\pm 50$ MeV has been extracted from the symmetry energy obtained by
the MDI interaction with the $x$ parameter between $0$ and $-1$. The
constraint $K_{\mathrm{asy}}=-500\pm 50$ MeV is quite consistent with the
very recent constraint of $K_{\mathrm{asy}}\approx -500_{-100}^{+125}$ MeV
from the study of neutron skin of finite nuclei \cite{Cen09}. Furthermore,
in the MDI interaction, we have $-311$ MeV $\leq K_{\mathrm{sat,2}}\leq -316$
MeV from the prediction of the MDI interaction with the $x$ parameter
between $0$ and $-1$. Therefore, for the MDI interaction, the magnitude of $%
K_{\mathrm{sat,2}}$ is significantly smaller than that of $K_{\mathrm{asy}}$
and is quite insensitive to the density dependence of the symmetry energy.
These features indicate that the high-order $J_{0}$ contribution to $K_{%
\mathrm{sat,2}}$ generally cannot be neglected.

\section{Summary and conclusions}

\label{summary}

We have constructed a phenomenological momentum-independent MID model which
can reasonably describe the general properties of symmetric nuclear matter
and the symmetry energy predicted by both the sophisticated isospin and
momentum dependent MDI model and the SHF approach with different Skyrme
forces. In particular, the density functional of the symmetry energy
constructed in the MID model is shown to be very flexible and can mimic very
different density behaviors by varying only one parameter.

Based on the MID model, we have studied in detail the second-order isospin
coefficient $K_{\mathrm{sat,2}}$ which is determined uniquely by $L$, $K_{%
\mathrm{sym}}$ and $J_{0}/K_{0}$. Our results indicate that the high-order $%
J_{0}$ contribution to $K_{\mathrm{sat,2}}$ generally cannot be
neglected, especially for larger $L$ values. In addition,
interestingly, it is found that there exists a nicely linear
correlation between $K_{\mathrm{sym}}$ and $L$ as well as between
$J_{0}/K_{0}$ and $K_{0}$ for the three different models used here,
i.e., the MDI interaction, the MID interaction, and the SHF approach
with $63$ Skyrme forces. From the MID model, the correlation between $K_{%
\mathrm{sym}}$ and $L$ is further shown to depend significantly on the value
of $E_{\text{\textrm{sym}}}(\rho _{0})$. These correlations and features
enable us to extract the values of the $J_{0}$ parameter and the $K_{%
\mathrm{sym}}$ parameter from the empirical information on $K_{0}$, $L$ and $%
E_{\text{\textrm{sym}}}(\rho _{0})$. In particular, using the empirical
constraints of $K_{0}=240\pm 20$ MeV, $25$ MeV $\leq E_{\text{\textrm{sym}}%
}(\rho _{0})\leq 35$ MeV,\ and $46$ MeV $\leq L\leq 111$ MeV in the
MID model leads to an estimate of $-477$ MeV $\leq
K_{\mathrm{sat,2}}\leq -241$ MeV.

While the estimated value of $-477$ MeV $\leq K_{\mathrm{sat,2}}\leq -241$
MeV in the present work has a small overlap with the constraint of $K_{\tau
}=-550\pm 100$\ MeV obtained in Ref. \cite{LiT07,Gar07} from recent
measurements of the isotopic dependence of the GMR in even-A Sn isotopes,
the magnitude of the constrained $K_{\tau }$ is still significantly larger
than that of $-477$ MeV $\leq K_{\mathrm{sat,2}}\leq -241$ MeV. Recently,
there are several works \cite{Sag07,Pie09} on extracting the value of the $%
K_{\mathrm{sat,2}}$ parameter based on the idea initiated by Blaizot
and collaborators that the values of both $K_{0}$ and
$K_{\mathrm{sat,2}}$ should be extracted from the same consistent
theoretical model that successfully reproduces the experimental GMR
energies of a variety of nuclei. These studies show that there is no
a single model (interaction) which can simultaneously describe
correctly the recent measurements of the isotopic dependence of the
GMR in even-A Sn isotopes and the GMR data of nuclei $^{90} $Zr and
$^{208}$Pb, which makes it difficult to accurately determine the
value of $K_{\mathrm{sat,2}}$ from the experimental GMR energies of
a variety of finite nuclei. As pointed out in \cite{Pie09}, these
features
seem to suggest that the $K_{\tau }=-550\pm 100$\ MeV obtained in Ref. \cite%
{LiT07,Gar07} may suffer from the same ambiguities already
encountered in earlier attempts \cite{Shl93} to extract the $K_{0}$
and $K_{\mathrm{sat,2}}$ of infinite matter from finite-nuclei
extrapolations. This problem remains as an open challenge, and both
experimental and theoretical insights are needed in the future.

\begin{acknowledgments}
The author thanks Professor En-Guang Zhao for helpful discussions
and encouragements. This work is supported in part by the National
Natural Science Foundation of China under Grants Nos. 10575071 and
10675082, MOE of China under project NCET-05-0392, Shanghai
Rising-Star Program under Grant No. 06QA14024, the SRF for ROCS, SEM
of China, the National Basic Research Program of China (973 Program)
under Contract No. 2007CB815004.
\end{acknowledgments}

\end{document}